# Feasibility considerations of a soft-x-ray distributed feedback laser pumped by an X-FEL

**Jean-Michel André[1,2], Karine Le Guen[1,2], Philippe Jonnard[1,2]**

[1] *Sorbonne Universités, UPMC Univ Paris 06, Laboratoire de Chimie Physique-Matière et Rayonnement, 11 rue Pierre et Marie Curie, F-75231 Paris cedex 05, France*

[2] *CNRS UMR 7614, Laboratoire de Chimie Physique-Matière et Rayonnement, 11 rue Pierre et Marie Curie, F-75231 Paris cedex 05, France*

**Abstract**

We discuss the feasibility of a soft-x-ray distributed feedback laser (DFL) pumped by an x-ray free electron laser (X-FEL). The DFL under consideration is a Mg/SiC bi-layered Bragg reflector pumped by a single X-FEL bunch at 57.4 eV, stimulating the Mg $L_{2,3}$ emission at 49 eV corresponding to the 3s-3d -> $2p_{1/2,3/2}$ transition. Based on a model developed by Yariv and Yeh and an extended coupled-wave theory, we show that it would be possible to obtain a threshold gain compatible with the pumping provided by available X-FEL facilities.

# 1. Introduction

Achieving a soft-x-ray laser with a well-defined photon energy, a narrow bandwidth and without spectral jitter remains a challenge at the present time. X-ray free electron lasers (X-FEL) have provided a huge step towards this goal, but some problems remain with this kind of laser, in particular concerning the spectral jitter and the spectral purity. This has been partially solved by the new facilities using the seeded self-amplification of spontaneous emission scheme, such as the FERMI facility. However, solid-state laser with optical resonator continues to be an interesting way for x-ray lasing as they are naturally free of spectral jitter.

In the soft-x-ray domain, there are two main obstacles in the way of lasing when using the scheme of stimulated radiation from an active solid-state medium: the first involves the pumping required to achieve the critical inversion and the second is the optical feedback. Considerations of the feasibility of x-ray laser [1] show that lasing in the conventional geometry, that is cavity with external reflectors, as proposed by different authors [2,3], requires considerable pump power. Using the distributed feedback provided by a Bragg reflector seems to be the preferred method [4–6]. At the present time, the availability of both X-FEL and efficient artificial Bragg reflectors should allow one to circumvent the problems of inversion and feedback.

The high flux available in each FEL pulse enables to achieve the population inversion : typical softx-ray FELs can deliver $10^{13}$ photons per shot, which is enough to make stimulated emission the dominating effect with respect to the spontaneous emission and the non-radiative decay channels (Auger process). The critical flux for inversion is estimated to be $10^{11}$ photons per shot. Recently M. Beye *et al.* [7] have demonstrated using the FLASH FEL facility, that under appropriate conditions, stimulated emission from crystalline silicon can be produced around 90 eV corresponding to the Si $L_{2,3}$ characteristic emission. Previously stimulation of emission from a single fluorescence line in a rare gas was demonstrated [8]. These successful experiments reinforce the idea that FEL radiation is appropriate for pumping softx-ray (solid-state) laser.

Periodic multilayers, developed since the eighties, should offer means to achieve the optical feedback in the soft-x-ray regime. It is now possible to fabricate efficient multilayers with periodic lattice spacing *d* suited to satisfy the Bragg condition in the softx-ray range. Multilayer optical cavity has many advantages : it provides a distributed feedback reducing the required pumping power [4–6]; it gives a shorter transit time than a standard cavity with external mirrors and eliminates the problem of alignment of multiple elements [9,10]. The spontaneous emission produced within a multilayer and diffracted under the Bragg condition has been recently observed using synchrotron radiation [11] : the synchrotron radiation from BEAR beamline (ELETTRA) excited the Mg Kα and Co Lα lines in Mg/Co multilayers which were Bragg-diffracted by the lattice. From these results, one expects that stimulated radiation can, in a same way, be Bragg-diffracted by the multilayer lattice. Thus, in principle, the two above mentioned obstacles can be overcome and the door seems open to consider the feasibility of DFL laser. That is what we propose to do in this work.

## 2. Our model of softx-ray distributed feedback laser

In this paper, in order to discuss the feasibility of a softx-ray distributed feedback laser (DFL) with the means at disposal atthe present time, we consider the following practical case. The pumping is achieved by an X-FEL offering the features of FERMI facility [12]while the optical feedback is provided by a Mg/SiC bi-layered Bragg reflector (see Figure 1). A single X-FEL bunch of the $12^{th}$ harmonic at 57.4 eVcreates holes in the $2p_{1/2}$and $2p_{3/2}$innershellsof Mg atoms (whose binding energies are 49.6 and 49.2 eV respectively) inthe corresponding Mg layers; the pumping is so intense ($10^{13}$ photons/bunch) that inversion of population occursin a way similar to the one reported in Ref. [7] and gives rise to the stimulatedMg $L_{2,3}$emission bandat 49 eV corresponding to the 3s-3d -> $2p_{1/2,3/2}$ transition.This emission band is quite large for solid Mg, about 6 eV, because it describes the density of occupied valence states.This large width possibly have an influence on the threshold inversion and gain but will not be considered. This band could also give rise to a small spectral tunability of the laser while the ultimate performance of the laser in terms of spectral linewidth should be given by the Schawlow-Townes formula [13]

In the stack, the SiC layers are passive and absorbing, and their losses need to be overcome by the gain of the active Mg layers according to the principle of the DFL. This problem will beaddressed in the following parts of this paper.

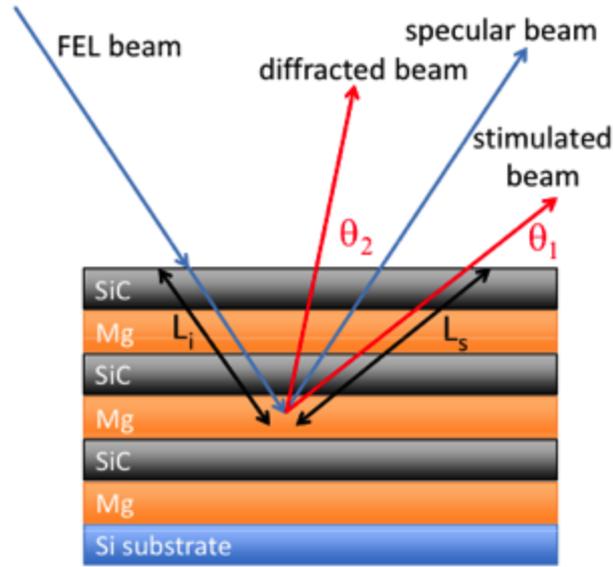

Fig 1: General scheme of the proposed experiment. Here the Bragg outgoing angle $\theta_2$ is different from the $\theta_1$ angle corresponding to the empirical condition given by Beye *et al.* [7].

Mg/SiC periodic structure acts as a Bragg reflector to ensure the feedback ; it is designed so that the Bragg outgoing angle $\theta_2$ (see Figure 1) for the Mg $L_{2,3}$ emission,coincides with the optimal outgoing angle $\theta_1$ for stimulated emission. This angle $\theta_1$ can be found from the empirical considerations given in Ref. [7] ; it is given by Arcsin($L_i/L_s$) where $L_i$ and $L_s$ are the attenuation lengths of the incident and stimulated radiations, respectively. For a $Mg_x(SiC)_{1-x}$ compound (having the same composition of the designed multilayer) with x=0.5, the angle $\theta_1$ is around 21.7°. The corresponding Bragg angle $\theta_2$ (equal to $\theta_1$)is achieved by a *d* spacing (bilayer thickness) equal to 35 nm and a $\gamma$ ratio equal to 0.53 (Mg layer thickness = $\gamma d$ = 18.55 nm and SiC thickness = (1- $\gamma$)d = 16.45 nm). This goal is not problematic since the angular reflection bandwidth of such a multilayer is large enough (around 7° for 10 bilayers) to include an uncertainty of a few degrees on the $\theta_2$ value. Moreover a peak reflectivity close to 67% is expected at 49 eV for an ideal structure.The spectral bandwith is estimated to be around 7 eV. In practice one should avoid that the Bragg angle $\theta_2$ coincides with the incoming angle of the X-FEL pumping radiation, because the detection of the lasing radiation should be affected by the specularly reflected X-FEL radiation.

The choice of the materials is motivated by the following considerations :
- Mg/SiCmultilayers have alreadybeen fabricated and they present a relatively high efficiency in the spectral domain of interest [14];

- The elements (Si, C) of the absorbing passive layers do not give rise to emission lines upon an excitation at 57.5 eV and their absorption coefficients are rather low at 49 eV.

The periodic multilayer can be described as a stack of bilayersmade up with a layer 1 (SiC) which is passive from the lasing point of view (absorbing medium with no gain inside) and with a layer 2 (Mg) which is active (having an effective gain resulting from the occurrence of the stimulated emission). Each type of layer is characterized by a complex optical index:

$$N^*(\omega) = N(\omega) + i\,\alpha(\omega)$$

(1)

Let us recall that in case of passive absorbing media, the imaginary term $\alpha(\omega) = k(\omega)$ is related to the linear absorbing coefficient $\mu(\omega)$ by the relationship:

$$k = \frac{\mu\,\lambda}{4\,\pi} = \frac{\mu\,c}{2\,\omega}$$

(2)

In the previous formula, ω stands for the radiation angular frequency related to the photon energy E by the Planck law, λ is the vacuum wavelength of the radiation and $c$ is the vacuum light velocity. The complex value of $N^*(\omega)$ for the absorbing media can be found in Ref. [15]. For active (amplifying) media, the material gain coefficient $g(\omega)$ will appear in the imaginary part of the complex optical index; thus in active absorbing media the sign of the imaginary part of $N^*(\omega)$ can be opposite to the sign of the imaginary part in passive absorbing media, depending on the sign of the net gain $\alpha(\omega) = g(\omega) - k(\omega)$.

The problem of deterioration of the multilayer stack under FEL irradiation has to be considered but we expect the Mg/SiC multilayer to withstand FEL irradiation or at least to minimize its damages. Indeed, as pointed in Ref. [7], one can expect only a few damagesin the Mg layers because the number of Auger processes, which are responsible of most of the damages, is substantially decreased since the number of core holes is reduced by the stimulated emission. Moreover, owing to the periodicity of the structure, close to the Bragg conditions, the anti-nodes of the electric field are located within the SiC layers, thus limiting the energy deposit by the stimulated beam. To minimize the energy deposited by the incident beam, the sample can be oriented so that the standing wave taking place within the stack is also so that the anti-nodes of the electric field are located within the SiC layers.Moreover, from thermodynamics considerations it has been shown that the duration of the optical properties of the various layers and thus the periodicity of the stack can exceed the fluorescence lifetime [10]. So, even if some damage should occur, its time scale would be longer than the one of the x-ray pulse [16–18].

## 3.Stimulated emission with a X-FEL pumping: atomic physics considerations

First, let us consider the lasing process from the point of view of atomicphysics as previously done by Yariv [1], without consideration ofthe optical feedback. Let us estimate the inversion density at which the gain $g$ becomes equal to the lossin a bulk Mg medium. The Mg absorption coefficient $\mu$(cm$^{-1}$) at the exciting X-FEL photon energy (57.4 eV) is :

$$\mu = \tau N$$

(3)

$N$ being the atomic density in the medium (at/cm$^3$) and $\tau$ the atomic absorption cross-section (345 10$^{-20}$ cm$^2$/at [15]). The density $N$ and the inversion density $\Delta N$ (density of atoms in the inverted state minus the density in the absorbing one)at which the gain $g$ and the loss $\mu$ compensate ($\mu = g$)satisfy:

$$\tau N = \Delta N \frac{e^2}{4\,\Delta v\, c\, m\, \epsilon_0}$$

(4)

the right-hand side term being the gain as given in Ref. [19];$e$ and $m$ the charge and the mass of electron respectively, $\epsilon_0$ the vacuum permittivity and $\Delta v$ the transition frequency line width. Numerically in SI units, one gets:

$$\frac{\Delta N}{N} = \tau \frac{4\, c\, m\, \epsilon_0}{e^2} \Delta v = 1.3\, 10^{-16} \Delta v$$

(5)

Taking into account the width of the L$_3$ level of Mg (0.03 eV [20]), $\Delta v$ is estimated to be 7.3 10$^{12}$ Hz so that :

$$\frac{\Delta N}{N} \approx 10^{-3}$$

(6)

This value of the fractional inversion density represents the minimum inversion to overcome the losses. Then it is possible to estimate the magnitude of pumping. Under steady state, the pumping surface power density $P_{pump}$ can be estimated to be :

$$P_{pump} = \frac{N_{ex}}{N} E_{pump} \frac{\omega_{ex}}{\sigma}$$

(7)

where $N_{ex}$ is the threshold density of excited atoms, $\sigma$ the absorption cross-section per atom at the pumping photon energy $E_{pump}$ and $\omega_{ex}$ the total relaxation rate of the laser (excited) level given by the transition line width. For our case involving the L$_3$ level of Mg excited with an X-FEL photon energy equal to 57.4 eV, from Eq. (7), $P_{pump}$ is given by:

$$P_{pump} \cong \frac{\Delta N}{N} Epump \frac{\omega_{ex}}{\sigma} \approx 2\ 10^{10}\ W/cm^2$$

(8)

This value is lower than the surface peak power density expected at the FEL-1 beamline of FERMI facility which can be estimated to be around $7\ 10^{12}$ W/cm$^2$ (5 GW in a 300 µm spot) [21]. Let us finish this section by a consideration on the laser linewidth. Assuming the validity of the Schawlow-Townes formula, the ultimate spectral bandwidth would be around 5 eV assuming a spectral bandwith of the cavity equal to 7 eV and a power of the laser around $10^{-2}$ W.

## 4. Lasing energy and threshold gain from the Yariv-Yeh model

We now consider lasing in Mg layers within a Mg/SiC periodic multilayer medium forming a DFL. The problem of the determination of the threshold gain (ThG) of a DFL in a multilayer resonator has been considered by Yariv and Yeh (Y-Y approach) in Refs. [4] and [22]. They obtain the ThG value by calculating the coordinates of the poles of the reflecting power R of the structure in the $\alpha_2$-$\omega$ plane where $\alpha_2$ is the term $\alpha$ in the active layer 2, see Eq. (1). These authors assert (without a proof) that these coordinates give the lasing frequencies and the corresponding values of the ThG.

For a structure based on $N$ bi-layered unit cells, the reflectivity $R_N$ is given by [6,23,24]:

$$R_N(\omega) = \left\| \frac{C\ (\omega)U_{N-1}}{A\ (\omega)U_{N-1} - U_{N-2}} \right\|^2$$

(9)

where $U_N$ denotes the Chebychev polynomials and $A$ and $C$ are coefficients whose expressions can be found in the previous references. The poles correspond to the zeros (in terms of frequencies $\omega$) of the denominator in Eq. (9); obviously at these poles, $R_N$ tends towards infinity. We have computed $R_N$ for the structure under consideration in this paper and described in Section 1,and plotted it versus the photon energy E and the term α₂. The results are givenin Figure 2.

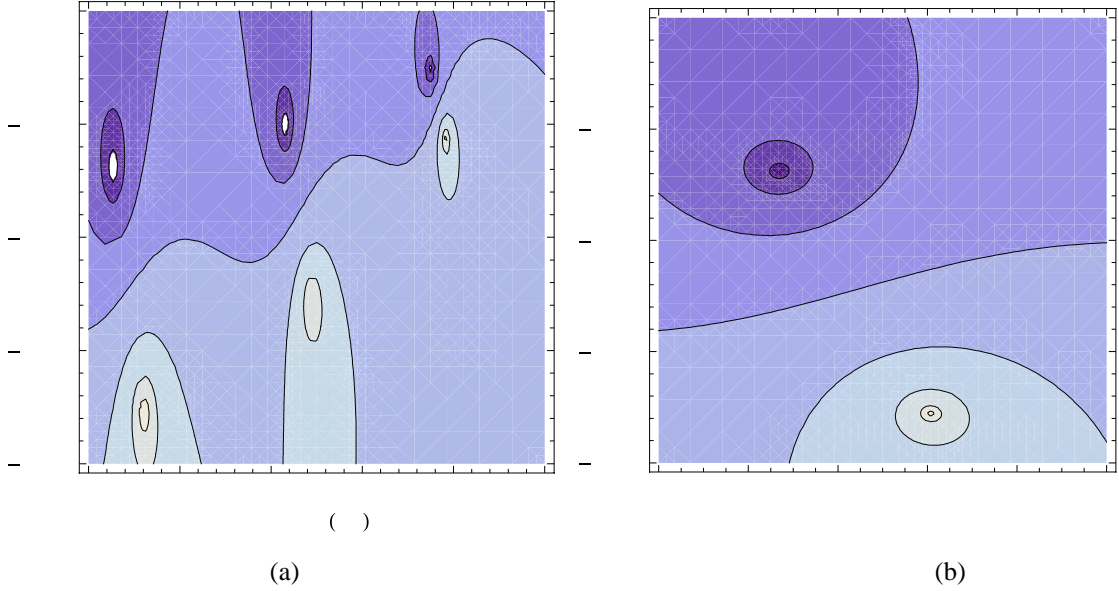

(a) (b)

Fig 2: (a) Reflectivity of the Mg/SiC multilayer described in Section 1 as a function of the photon energy and the net gain in the Mg layers. (b) Zoom, showing the location of the pole in reflectivity. The brighter the colour, the higher is the reflectivity.

Domains displayed in white area correspond to large values of $R_N$ and their center gives the lasing condition. Such a lasing condition is foundnear the first Bragg region at E = 45.61 eV with a ThG value α₂ = 17.8 10⁻³ that is a value of μ₂ = 78600 cm⁻¹.

## 5. Lasing energy and threshold gainfrom the coupled-wave theory

The coupled-wave (CW) theory was initially implemented by Kogelnik and Shank to determine both the resonant frequencies and gain threshold criteria [25]; they considered a symmetric device (valid for Bragg diffraction at angle near π/2) and didnot take into account reflections at theexternal boundaries. The model was extended by Chinn [26]to the case of non-zero reflectivity at the extremities of the lasing medium. In this work, we consider the general case ofa non-symmetric device and non-zero reflectivity at the facets. In this model, one assumes a spatial modulation (along the stack axis $z$)of the refractive index $n^*(\omega, z)$ and of the material gain $g(\omega, z)$ of the form:

$$n^*(\omega, z) = \bar{n}^*(\omega) + \Delta n^*(\omega, z) \cos(2\, \beta_B\, z)$$

(10)

$$g(\omega, z) = \bar{g}(\omega) + \Delta g(\omega, z) \cos(2\, \beta_B\, z + \psi_g)$$

(11)

where

$$\beta_B \stackrel{\text{def}}{=} \frac{\pi}{d} = \beta - \delta\beta \qquad (12)$$

Here the refractive index $n^*(\omega, z)$ includes the absorbing term $k(\omega)$ only in its imaginary part; in this sense it must be distinguished from the optical index $N^*(\omega)$, see Eq. (1). The quantity $\beta = n^* \beta_0 = n^* \frac{\omega}{c}$ is the propagation constant of a wave of frequency ω; the term $\delta\beta$ is called the detuning parameter. The length L of the structure is given by the product $Nd$ where N is the number of bilayers. In a general model, $\psi_g$ is the dephasing between the index $n^*(\omega, z)$ and the amplitude gain constant $g(\omega, z)$ profile; in our numerical simulations, we consider only a system with no dephasing, i.e. $\psi_g = 0$.

In the CW model, one introduces the coupling coefficients $\kappa^+$ and $\kappa^-$ given by:

$$\kappa^+ = \kappa - i\,\kappa_g\, e^{i\psi_g}$$
$$\kappa^- = -\kappa + i\,\kappa_g\, e^{i-\psi_g} \qquad (13)$$

with the coupling constant

$$\kappa = \beta_0 \Delta n^* / 2 \qquad (14)$$

and

$$\kappa_g = \Delta g / 4 \qquad (15)$$

The coupling taking into account the gain is described by the quantity:

$$\Delta\beta = \delta\beta - i\frac{g}{2} \qquad (16)$$

According to the principle of the CW model, one considers that two counter-waves R and S are coupled by backward Bragg scattering and the electric field within the structure, is written:

$$E(z) = R(z)e^{-i\beta_B z} + S(z)e^{+i\beta_B z} \qquad (17)$$

The calculations giving the expressions of *R(z), S(z)*, the resonant frequencies (lasing photon energies) and the gain threshold are detailed in Appendix A. The lasing condition, that is the gain threshold, is given by :

$$\frac{(\Gamma^+ - \rho_L e^{2i\beta_B L})}{(1 - \Gamma^- \rho_L e^{2i\beta_B L})} \frac{(\Gamma^- - \rho_0)}{(1 - \rho_0 \Gamma^+)} = e^{-2\gamma L}$$

(18)

where γ is an eigenvalue given as a solution of the eigenequation of the coupled-wave system of equations (see Appendix A) :

$$\gamma = \pm i\sqrt{\Delta\beta^2 - \kappa^- \kappa^+}$$

(19)

$\rho_0$ and $\rho_L$ are the reflection coefficient at $z = 0$ and $z = L$ respectively and the terms $\Gamma^+, \Gamma^-$ are related to the coupling coefficients $\kappa^+$ and $\kappa^-$, to the detuning parameter $\delta\beta$ and to the eigenvalues γ by :

$$\Gamma^- \stackrel{def}{=} \frac{R^-}{S^-} = \frac{-\kappa^+}{i\gamma + \Delta\beta}$$

(20)

and

$$\Gamma^+ \stackrel{def}{=} \frac{S^+}{R^+} = \frac{\kappa^-}{i\gamma + \Delta\beta}$$

(21)

The eigenvalues γ can be found numerically by combining Eqs. (18) and (19). Let us note that Eq. (18) reduces to:

$$\Gamma^+ \Gamma^- = e^{-2\gamma L}$$

(22)

for the case of vanishing reflection at the boundaries, that is:

$$\frac{-\kappa^- \kappa^+}{(i\gamma + \Delta\beta)^2} = e^{-2\gamma L}$$

(23)

which by using Eq. (19), leads to:

$$\gamma = i\Delta\beta \, \text{th} \, \gamma L$$

(24)

that is the formula (19) given by Kogelnik and Shank in [25]. Always in the case of vanishing reflection at the boundaries, by combining Eqs. (19) and (24), it follows that the eigenvalues γ satisfy the transcendental equation :

$$\gamma = \pm i \sqrt{\kappa^- \kappa^+} \sinh \gamma L$$

(25)

similar to formula(18) in Ref. [25]. Let us emphazise that the assumption of vanishing reflection at the boundaries holds in the x-ray regime above the critical angle. By solving numerically Eqs. (18) and (19), we found a value of the threshold gain $\alpha_2$ around 20 $10^{-3}$ at a lasing energy of 42 eV in fair agreement with the value obtained from the Y-Y mode, 18 $10^{-3}$.

It is valuable to note that the lasing condition given by Eq. (18) is equivalent to find the poles of the reflection or transmission coefficientsobtained in the framework ofthe CW theory asshown in Appendix B.The structure acts as an oscillator since it gives finite output fields without input. This can be considered as a justification of the Y-Y approach implemented in Section 4.

## 6. Conclusions

From atomic physics considerations, it appears that the power density delivered by available X-FELs such as FERMI is sufficient to generate stimulated emission in a low-Z element such as Mg, in agreement with the experimental results obtained by Beye *et al.* [7]. In the DFL scheme,the threshold gain deduced from the Y-Y model or the CW theory, that amounts to be around 20 $10^3$ cm$^{-1}$.

The calculations presented in this work show the feasibility of a softx-ray DFL based on superlattices such as the Mg/SiC system under consideration and excited by an X-FEL having the features of FERMI. More elaborated systems with non-uniform gain and coupling [27] should make it possible to lower the threshold gain. This case will be studied in a forthcoming paper.

The present calculations were performed for ultra-soft x-ray incident and stimulated radiations. Let us note that there exist now (or are under construction) some X-FEL sources whose photon energy can be as high as 10 keV (0.12 nm). Under these conditions it would be possible to use a single crystal as the optical cavity, as previously envisaged [5,10,28], to obtain aDFL laser working in a harder x-ray range. For example, instead of making the Beye's experiment with the Si $L_{2,3}$ characteristic emission, it could be possible to consider to use the Si K$\alpha$ emission (2p – 1s transition). Its energy is 1740 eV (0.71 nm) and the Si 1s binding energy is 1840 eV (0.67 nm). Thus, by using an incident photon energy equal to or larger than the Si 1s binding energy and a Si crystal whose planes parallel to the surface are (110)and have a*2d* equal to 0.77 nm, it would be possible to detect the Si K$\alpha$stimulated emission in the Bragg condition near a glancing angle of 68°, whereas the angle for stimulated emission given by the criterion on the attenuation lengths [7] leads to an angle of 6°.

## Acknowledgments

Dr. Emiliano Principi from FERMI @ ELETTRA is thanked for valuable discussions.

# Appendix A

Introducing Eq. (11) in the scalar wave equation and neglecting the second spatial derivatives according to the principle of the CW model, one obtains for $R(z)$ and $S(z)$ the coupled linear system of equations :

$$\begin{pmatrix} \dot{R}(z) \\ \dot{S}(z) \end{pmatrix} = \begin{pmatrix} \Delta\beta & \kappa^+ \\ \kappa^- & -\Delta\beta \end{pmatrix} \begin{pmatrix} R(z) \\ S(z) \end{pmatrix}$$

(A.1)

where the dot stands for the derivative with respect to $z$. The general solution of the system (A.1) can be expressed from the two eigenmodes,

$$\begin{pmatrix} R(z) \\ S(z) \end{pmatrix} = \begin{bmatrix} R^+ \\ S^+ \end{bmatrix} e^{\gamma z} + \begin{bmatrix} R^- \\ S^- \end{bmatrix} e^{-\gamma z}$$

(A.2)

Introducing the reflection coefficients $\Gamma^-$ given by Eqs. (14) and (15) respectively, we get:

$$\begin{pmatrix} R(z) \\ S(z) \end{pmatrix} = R^+ \begin{bmatrix} 1 \\ \Gamma^+ \end{bmatrix} e^{\gamma z} + S^- \begin{bmatrix} \Gamma^- \\ 1 \end{bmatrix} e^{-\gamma z}$$

(A.3)

Using Eq. (A.2) at $z = 0$ gives:

$$\begin{pmatrix} R(0) \\ S(0) \end{pmatrix} = \begin{pmatrix} 1 & \Gamma^- \\ \Gamma^+ & 1 \end{pmatrix} \begin{pmatrix} R^+ \\ S^- \end{pmatrix}$$

(A.4)

It is then possible to relate the field at each boundary:

$$\begin{pmatrix} R(L) \\ S(L) \end{pmatrix} = [P] \begin{pmatrix} R(0) \\ S(0) \end{pmatrix}$$

(A.5)

with

$$P_{11} = \frac{e^{\gamma L} - e^{-\gamma L}\Gamma^-\Gamma^+}{1 - \Gamma^-\Gamma^+}$$

$$P_{12} = \frac{\Gamma^-(e^{-\gamma L} - e^{\gamma L})}{1 - \Gamma^-\Gamma^+}$$

$$P_{21} = \frac{\Gamma^+(e^{+\gamma L} - e^{-\gamma L})}{1 - \Gamma^-\Gamma^+}$$

$$P_{22} = \frac{e^{-\gamma L} - e^{+\gamma L}\Gamma^-\Gamma^+}{1 - \Gamma^-\Gamma^+}$$

(A.6)

One considers now that some non-vanishing reflection exists at the boundaries $z = 0$ and $z = L$. The boundary conditions are :

i/ At $z = L$,

$$S(L)e^{-i\beta_B L} = \rho_L R(L)e^{i\beta_B L}$$

(A.7)

that is, taking into account Eqs. (A2-A5)

$$R^+\left(\rho_L e^{2 i\beta_B L} - \Gamma^-\right)e^{2\gamma L} - S^-\left(1 - \Gamma^- e^{2 i\beta_B L}\right) = 0$$

(A.8)

where $\rho_L$ is the reflection coefficient at $z = L$.

ii/ At $z = 0$,

$$R(0) = \rho_0 S(0)$$

(A.9)

that is,

$$R^+(1 - \rho_0 \Gamma^+) + S^-(\Gamma^- - \rho_0) = 0$$

(A.11)

where $\rho_0$ is the reflection coefficient at $z = 0$.

The lasing condition can be obtained by considering that there are non-trivial solutions of the system of equations $\Sigma$ formed by Eqs. (A.9) and (A.10) if

$$Det[\Sigma] = 0$$

(A.11)

The lasing condition given by Eq. (A.11) can be written explicitly as follows :

$$\frac{\left(\Gamma^+ - \rho_L e^{2 i\beta_B L}\right)}{\left(1 - \Gamma^- e^{2 i\beta_B L}\right)} \frac{(\Gamma^- - \rho_0)}{(1 - \rho_0 \Gamma^+)} = e^{-2\gamma L}$$

(A.12)

From Eqs. (A.2), (14) and (15)

$$\begin{pmatrix} R(z) \\ S(z) \end{pmatrix} = \begin{bmatrix} 1 \\ \frac{1}{\Gamma^+} \end{bmatrix} R^+ e^{\gamma z} + \begin{bmatrix} \Gamma^- \\ 1 \end{bmatrix} S^- e^{-\gamma z}$$

(A.13)

Moreover, by inverting Eq. (A.3) and taking into account Eq. (A.9), we get:

$$\begin{pmatrix} R^+ \\ S^- \end{pmatrix} = \frac{S(0)}{1 - \Gamma^-\Gamma^+} \begin{pmatrix} 1 & -\Gamma^- \\ -\Gamma^+ & 1 \end{pmatrix} \begin{pmatrix} \rho_0 \\ 1 \end{pmatrix}$$

(A.14)

Inserting Eq. (A.14) in Eq. (A.13) gives the following expressions for the field $R(z)$ and $S(z)$ after dropping an arbitrary amplitude factor:

$$R(z) = \frac{1}{1 - \Gamma^-\Gamma^+}[(\rho_0 - \Gamma^-)e^{\gamma z} + \Gamma^-(1 - \rho_0 \Gamma^+)e^{-\gamma z}]$$

$$S(z) = \frac{1}{1 - \Gamma^-\Gamma^+}\left[\frac{1}{\Gamma^+}(\rho_0 - \Gamma^-)e^{\gamma z} + (1 - \rho_0\Gamma^+)e^{-\gamma z}\right]$$

(A.15)

For zero-reflectivity at $z = 0$

$$R(z) = \frac{\Gamma^-}{(\Gamma^-\Gamma^+ - 1)}[e^{\gamma z} - e^{-\gamma z}]$$

$$S(z) = \frac{1}{\Gamma^+(\Gamma^-\Gamma^+ - 1)}[\Gamma^- e^{\gamma z} - \Gamma^+ e^{-\gamma z}]$$

## Appendix B

The Fresnel formula in matrix form [24] leads to :

$$\begin{pmatrix}R^+\\S^-\end{pmatrix}_{z=L} = \frac{1}{\sqrt{1 - |\rho_L|}}\begin{pmatrix}1 & -\rho_L\\-\rho_L & 1\end{pmatrix}\begin{pmatrix}R(L)e^{i\beta_B L}\\S(L)e^{-i\beta_B L}\end{pmatrix}$$

(B.1)

and

$$\begin{pmatrix}R(0)\\S(0)\end{pmatrix} = \frac{1}{\sqrt{1 - |\rho_0|}}\begin{pmatrix}1 & \rho_0\\\rho_0 & 1\end{pmatrix}\begin{pmatrix}R^+\\S^-\end{pmatrix}_{z=0}$$

(B.2)

Using Eqs. (B.1) and (B.2), it comes :

$$\begin{pmatrix}R^+\\S^-\end{pmatrix}_{z=L} = [G]\begin{pmatrix}R^+\\S^-\end{pmatrix}_{z=0}$$

(B.3)

where

$$[G] = \frac{1}{\sqrt{1 - |\rho_L|}}\frac{1}{\sqrt{1 - |\rho_0|}}\begin{pmatrix}1 & -\rho_L\\-\rho_L & 1\end{pmatrix}\begin{pmatrix}e^{i\beta_B L} & 0\\0 & e^{-i\beta_B L}\end{pmatrix}[P]\begin{pmatrix}1 & \rho_0\\\rho_0 & 1\end{pmatrix}$$

(B.4)

Consequently, the global transmission coefficient $\tau$ is:

$$\tau \stackrel{\text{def}}{=} \left.\frac{R^+_{z=L}}{R^+_{z=0}}\right|_{S^-_{z=L}=0} = \frac{Det[G]}{G_{22}}$$

(B.5)

while the global reflection coefficient $\rho$ is :

$$\rho \stackrel{\text{def}}{=} \left.\frac{S^-_{z=0}}{R^+_{z=0}}\right|_{S^-_{z=L}=0} = \frac{-G_{21}}{G_{22}}$$

(B.6)

The condition $G_{22} = 0$ is identical to Eq. (A.12).